\documentclass[prb,twocolumn,floatfix,showpacs,superscriptaddress]{revtex4}

\usepackage{graphicx}
\usepackage{amsmath,amssymb}
\usepackage{psfrag}
\usepackage[usenames]{color}

\newcommand{\bR}{\mathbf{R}}

\begin{document}
\title{Biexciton recombination rates in self-assembled  quantum dots}
\author{Michael Wimmer}
\email{Michael.Wimmer@physik.uni-regensburg.de}
\affiliation{Institut f\"ur Theoretische Physik,
Universit\"at Regensburg, 93040 Regensburg, Germany}
\affiliation{Department of Physics and Astronomy,
Arizona State University, Tempe AZ 85287-1504}
\author{S. V. Nair}
\affiliation{Centre for Advanced Nanotechnology, University of Toronto,
Toronto ON M2J 1K2, Canada}
\author{J. Shumway}
\email{john.shumway@asu.edu}
\affiliation{Department of Physics and Astronomy,
Arizona State University, Tempe AZ 85287-1504}
\date{\today}
\newcommand{\re}[0]{{\mathbf r}^{\text{e}}}
\newcommand{\rh}[0]{{\mathbf r}^{\text{h}}}
\newcommand{\rep}[0]{{\mathbf r}^{\text{e}'}}
\newcommand{\rhp}[0]{{\mathbf r}^{\text{h}'}}
\newcommand{\RN}[0]{{\mathbf R}_N}
\newcommand{\RNm}[0]{{\mathbf R}_{N-1}}
\begin{abstract}
The radiative recombination rates of interacting electron-hole pairs 
in a quantum dot are strongly affected by quantum correlations among 
electrons and holes in the dot.
Recent measurements of the biexciton recombination rate in single 
self-assembled quantum dots have found values spanning from two times 
the single exciton recombination rate to values well below the exciton 
decay rate. In this paper, a Feynman path-integral formulation is 
developed to calculate recombination rates including thermal 
and many-body effects.
Using real-space Monte Carlo integration, the path-integral
expressions for realistic three-dimensional models of InGaAs/GaAs, CdSe/ZnSe, 
and InP/InGaP dots are evaluated, including anisotropic 
effective masses. Depending on size, radiative rates of typical dots lie in the
regime between strong and intermediate confinement. The results
compare favorably to recent experiments and calculations on related
dot systems. Configuration interaction calculations using uncorrelated 
basis sets are found to be severely limited in calculating decay rates.
\end{abstract}


\maketitle

\section{Introduction}

The small size and strong optical properties of self-assembled quantum dots 
(QDs) make them appealing candidates for optoelectronic 
devices.\cite{Wojs:1996,Bimberg:1999}
When light is absorbed, photons create electron-hole (eh) pairs (excitons) 
that become confined in the quantum dot. Recent
photoluminescence (PL) spectra have measured the recombination energy
of electron-hole pairs with meV resolution.\cite{Dekel:2000,Santori:2002}
 Analysis of single dot PL spectra at different incident light intensities 
reveals that the exciton recombination energy is shifted by other ``spectator''
excitons and free charges in the dot.\cite{Dekel:2000,Santori:2002}
For example the recombination energy is 
red-shifted a few meV by the presence of a spectator 
exciton.\cite{Dekel:2000,Santori:2002}
 Detailed understanding of the effect of spectators on recombination
is important for non-linear optical applications, such as quantum logic gates 
\cite{Li:2003} or turnstiles.\cite{Kim:1999,Zrenner:2002}

The rates of the PL processes determine the steady-state occupation of the 
dots for a given incident intensity.\cite{Santori:2002}
 Time-resolved photoluminescence measurements
can track the electron-hole recombination rate in single self-assembled
quantum dots.\cite{Santori:2002} Recent experiments give 
differing results about
the decay rate of the biexciton relative to that of an isolated exciton in
the same dot.
Measurements on a single CdSe/ZnSe dot find a biexciton decay rate 
$\Gamma_{XX}$ about equal to the exciton rate $\Gamma_X$,\cite{Bacher:1999}
while other experiments on similar sized CdSe/ZnSe QDs report a biexciton
decay rate twice the exciton rate.\cite{Patton:2003}
Similar measurements in InGaAs have found 
$\Gamma_{XX}/\Gamma_X\!\approx\! 1.5$,\cite{Santori:2002} 
$\Gamma_{XX}/\Gamma_X\!\approx\! 2$,\cite{Thompson:2001,Ulrich:2005}
and even $\Gamma_{XX}/\Gamma_X\!\approx\! 0.33$.\cite{Kono:2005}
 
Theoretically, there are two limits to consider for recombination rates.
In the strong confinement limit, the exciton and biexciton wave function
is a simple product of the electron and hole single-particle 
wave functions in the dot. Coulomb interactions are assumed to
only slightly perturb the wave function. In that case the recombination
rates contain matrix elements of the single-particle wave functions,
which are the same for excitons and biexcitons.
Taking into account the number of allowed decay channels, the
biexciton should decay at twice the exciton rate, $\Gamma_{XX}/\Gamma_X=2$.
The other limit is the weak confinement limit, which applies when the 
exciton binding energy significantly exceeds the single-particle level 
spacing of the dot. In that limit, the exciton or biexciton is bulk-like, 
bound together as a small composite particle. This exciton or 
biexciton unit is weakly confined in a dot much
larger than the exciton or biexciton radius. In this case the dipole matrix
element is dominated by the exciton or biexciton structure, which is 
independent of dot size. The composite particle has a coherent 
wave function that extends across the volume of the dot, 
leading to constructive addition of radiative matrix
elements for exciton decay. Thus, in the weak confinement limit, 
the radiative decay rate of the exciton
{\em increases} with dot size, until the dot diameter approaches the wavelength
of the emitted light. For the biexciton, the exciton final 
state after recombination suppresses this constructive enhancement, 
significantly reducing the value of $\Gamma_{XX}/\Gamma_X$ 
in the weak confining limit. 
In the intermediate regime, the exciton wave function generally cannot 
be separated, except for some special choice of the external potential, 
such as a harmonic confinement.\cite{Sugawara:1995} Still, the coherent 
extent of the many-particle wave function---the 
\emph{coherence volume}\cite{Sugawara:1995}---leads to an increasing 
decay rate with increasing dot size and will play an important role in 
the interpretation of our results.
In this paper we show that the radiative decay rates of typical 
self-assembled dots lie in the regime between strong and intermediate 
confinement.

Theoretical descriptions of single particle (electron or hole) states 
in quantum dots have improved greatly in the last ten 
years, \cite{Bimberg:1999,Williamson:1999c,%
Heitz:2000,Santoprete:2003} yet the description
of exciton or multi-exciton states is not as well developed. The energies of
states with several electrons and holes are usually treated within first-order
perturbation theory. Some spectral energies, such as the biexciton shift,
require treatment of correlation with configuration interaction (CI) or 
quantum Monte Carlo (QMC) techniques.\cite{Shumway:2001a}
The limited accuracy of approximate CI wave functions
is known to affect the calculated energies.\cite{Shumway:2001a}

There have been a few attempts to calculate biexciton decay rates 
in quantum dots. Takagahara used a variational calculation to determine 
the decay rate of a biexciton in an infinite barrier
spherical dot with dielectric effects.\cite{Takagahara:1989}
More recently, Ungier {\em et al.}~have calculated the biexciton decay rate 
for zincblende and wurtzite structures.\cite{Ungier:1999} 
Using CI expansions, Corni {\em et al.}~have studied the size dependence of
exciton and biexciton recombination rates in strain-induced 
dots.\cite{Corni:2003} 
These dots are formed in a near-surface InGaAs/GaAs quantum well by the stress
field of an InP self-assembled island grown on the surface. These dots have 
much shallower confinement than self-assembled dots, are often much larger, 
and are well-approximated by truncated 2-d parabolic confinement. 
This puts the strain induced dots well into the weak confinement regime, 
contrary to the more common self-assembled dots that are subject of this paper.
Recently, Narvaez {\em et al.}~have performed CI calculations on 
InGaAs/GaAs self-assembled dots beyond the effective mass 
approximation using pseudopotentials.\cite{Narvaez:2005}
These CI results must be viewed with some caution
since decay rates are more sensitive than energies 
to errors in the wave function, as we will show in this paper. 

In this paper we develop a Feynman path integral description of exciton and 
biexciton recombination rates. This technique can be easily applied 
to complicated dot geometries, does not depend on a finite basis set, and fully
treats correlation. In Sec.~\ref{method} we derive a path integral expression 
for the recombination rate. This expression is then evaluated using a 
real-space Monte Carlo technique that we introduce in Sec.~\ref{compmethod}. 
In Sec.~\ref{dottest} we apply our path integral technique to a model 
system and compare with full CI calculations. In Sec.~\ref{results} 
we apply both the path integral technique 
and the CI expansion to realistic three-dimensional models of InGaAs/GaAs, 
CdSe/ZnSe, and InP/InGaP dots and compare to single dot experiments.
While our path integral method is currently restricted to 
single-band effective mass approximation (EMA) models, the insights,
trends, and even quantitative rates revealed in these 
make them quite useful, as we conclude in Sec.~\ref{conclusions}.

\section{Method}\label{method}

\begin{figure}
\centerline{\includegraphics[width=0.92\linewidth]{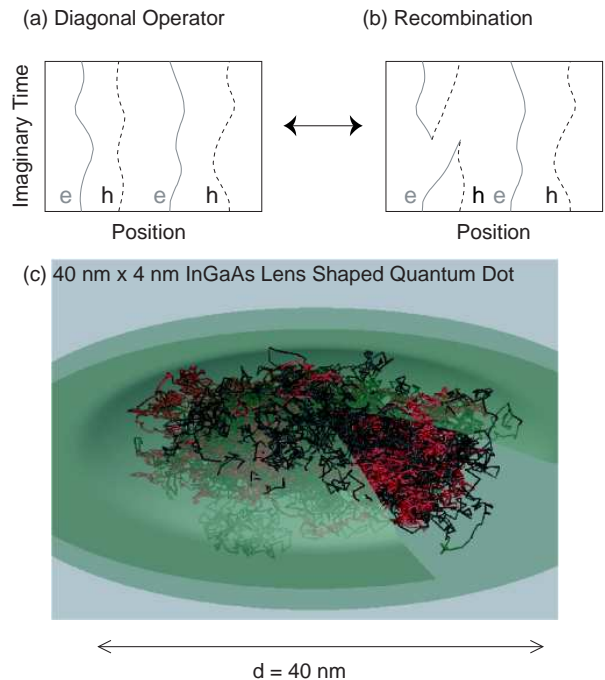}}
\caption{(Color online)
Illustration of our path-integral calculation of biexciton 
recombination rates in 
self-assembled quantum dots. We express the rate as a ratio
of path integrals with (a) diagonal and (b) radiating 
constraints, see Eq.~(\ref{eq:pirate}). We evaluate the path
integrals using Monte Carlo integration on realistic three-dimensional
models. A typical path contributing to the integrals for an InGaAs/GaAs
dot is shown in (c). These paths sample the probability density, energy, rate,
and other properties of the radiating states.
\label{fig:path}}
\end{figure}

Starting from a standard treatment of electron-hole
radiative recombination in the effective mass approximation, we
rewrite the square of the matrix element in the rate equation as
a path integral expression. In the path integral formalism,
we will show that
the rate is proportional to the ratio of two path integrals:
one with the standard thermal trace, 
and the other with a ``radiating'' configuration 
that pairs an electron and hole, as illustrated
in Fig.~\ref{fig:path}(a) and (b).

\subsection{Exciton recombination rate within the effective mass approximation}
Our Hamiltonian is a commonly used effective mass model, 
\begin{equation}
\label{eq:hamil}
\begin{split}
H &= \sum_{N_h} \left(\frac{\mathbf{p}^2_h}{2m^*_h} 
+ V_h(\mathbf{r}_h)\right)\\
&+ \sum_{N_e} \left(\frac{\mathbf{p}^2_e}{2m^*_e} 
+ V_e(\mathbf{r}_e)\right)
+ \frac{1}{2} \sum_{i\neq j} \frac{q_i q_j}{\epsilon r_{ij}},
\end{split}
\end{equation}
where $V_e$ and $V_h$ describe a lens-shaped confining potential and 
the hole effective masses $m_h$ are anisotropic. 
In contrast to previous approaches to this problem, we do not construct
single-particle or variational wave functions. Rather, we use Metropolis
Monte Carlo to sample the recombination rate directly from a
path integral.
The path integral Monte Carlo (PIMC) method allows us to calculate 
the density matrix for the Hamiltonian, Eq.~(\ref{eq:hamil}).
As we describe below, this is an essentially exact solution
without basis set problems or the difficulties of variational approaches.
A snapshot of a typical path for an electron-hole pair in our simulation is
shown in Fig.~\ref{fig:path}(c). A sum over all such paths is a
complete quantum mechanical solution for the model, Eq.~(\ref{eq:hamil}).

The rate of spontaneous decay of an exciton into a photon is the sum 
of the rates
of all possible decay processes. For generality, we consider a state 
$\Phi^\alpha_i$ with $N$ electron-hole pairs decaying to a state 
$\Phi^\alpha_f$ with $N-1$ pairs.
The rate of spontaneous emission into a photon with polarization 
$\hat{\lambda}$, momentum $\hbar\mathbf{k}$, and energy $\hbar\omega$,
in a medium with index of refraction $n$ $(\approx\sqrt\epsilon)$, is
\begin{equation}
\label{eq:goldenrule}
\frac{d\Gamma_{\mathbf k \hat{\lambda}}^\alpha}{d\Omega} =
\frac{n\omega e^2}{hc^3} |\langle\Phi^\alpha_f|
\mathbf{j}_{\mathbf k \hat{\lambda}} 
\cdot \hat{\lambda}|\Phi^\alpha_i\rangle|^2,
\end{equation}
which follows from Fermi's golden rule.
Since the emitted photon has energy slightly less than the band gap (typically
1-3 eV), the photon wavelength is much greater than the dimensions of the
self-assembled dot (typically 5-50 nm), so we take the $k\to 0$
limit in the current operator, $\mathbf j_{\mathbf k\to 0, \hat{\lambda}}$. The
usual approximation for the exciton decay rates in semiconductors is
to use the envelope approximation, in which the single particle
wave functions are approximated as an envelope times a periodic Bloch function,
$\phi(\mathbf r) = \psi(\mathbf r) u(\mathbf r)$. Then the current operator
splits into a delta function on the envelope and the current operator
$\mathbf j = \mathbf p/m$ on the Bloch function. The momentum
matrix element between the conduction band (CB) and valence band (VB) 
Bloch functions 
is given by the Kane parameter, $E_{\text{P}}=|\langle \text{CB}|
\mathbf{p}|\text{VB}\rangle|^2/2m$. 
Since all significant transitions occur in an energy range given by the 
coulomb interaction (a few tens of meV), which is much smaller than the
gap energy, we take the usual approximation $\hbar\omega \approx 
E_{\text{gap}}$.
Thus, within the envelope approximation, the recombination rate due to
transition $\alpha$ is approximately
\begin{equation}
\label{eq:envelopedipole}
\Gamma^\alpha
=\frac{2 n E_{\text{gap}} E_{\text{P}} e^2}{3 \hbar^2 c^3m} |I_N^\alpha|^2,
\end{equation}
where the point contact matrix element $I$ is the overlap integral of the 
initial and final envelope functions,
\begin{equation}
\begin{split}
\label{eq:pointcontact}
I_N^\alpha = \int &{\psi_N^\alpha}^*(\RN)\psi_{N-1}^\alpha(\RNm)\\
&\delta^3(\re_N-\rh_N)\,d^{3N}\RN .
\end{split}
\end{equation}

\subsection{Path integral expression for the rate}

The determination of exciton and biexciton recombination rates using 
Eqs.~(\ref{eq:envelopedipole}) and (\ref{eq:pointcontact}) faces
two difficulties. First, the initial and final states contain
several interacting particles, for which correlation must be treated carefully.
Second, the total rate is a sum over all possible transitions, $\Gamma
=\sum_\alpha\Gamma^\alpha$. These difficulties may be treated explicitly for 
model systems (such as harmonic oscillators), 
but generally the direct determination of the 
matrix elements and rates in a wave function
representation leads to approximations. For example, 
Takagahara's variational calculation\cite{Takagahara:1989} treats 
correlation very well for a single transition 
from a biexciton to the exciton ground state, but is limited to 
very symmetric, spherical 
QDs. On the other hand, Ungier 
{\em et al.}\cite{Ungier:1999}~treat structural details of the dot with
much care (even beyond the envelope function above),
but the correlation is only
partially included, an approximation known to underestimate biexciton
binding energies.\cite{Shumway:2001a} CI calculations
can in principle solve the full many-particle Schr\"odinger
equation for excitons and biexcitons
for EMA models\cite{Corni:2003} and pseudopotentials,\cite{Narvaez:2005}
but may be severely limited by the underlying basis set, as we will show.

We now derive a path-integral solution for the total recombination rate 
that will fully treat correlation, thermally distribute the initial states,
and include all final states. 
To relate Eqs.~(\ref{eq:envelopedipole}) and (\ref{eq:pointcontact}) to a 
path integral, we begin by squaring the point contact matrix element,
\begin{equation}
\begin{split}\label{eq:dmat}
|I_N^\alpha|^2 = &\iint \rho^\alpha_N(\RN,\RN')\rho^\alpha_{N-1}(\RNm',\RNm)\\
&\;\;\delta(\re_N-\rh_N)\delta(\rep_N-\rhp_N) d\RN\,d\RN',
\end{split}
\end{equation}
where $\rho^\alpha_N$ and $\rho^\alpha_{N-1}$ are the density matrices of the 
initial and final states. As in Ref.~\onlinecite{Dekel:2000}, we assume that 
the carriers reach thermal
equilibrium before the transition, and use the thermal density matrix
of $N$ electron-hole pairs, $\rho_N(\RN,\RN';\beta)$. 
The final state can take on any value, so we sum over all final states,
yielding $\rho_{N-1}(\RNm',\RNm)=\delta^{3(N-1)}(\RNm-\RNm')$.
After integrating out the $\RN'$ coordinates in Eq.~(\ref{eq:dmat}) with this 
delta function
and using Eq.~(\ref{eq:envelopedipole}), we find the temperature-dependent
radiative recombination rate,
\begin{equation}
\label{eq:pirate}
\Gamma_N(\beta) = \frac{2nE_{\text{gap}}E_{\text{P}}e^2}{3\hbar^2c^3m} 
\langle|I_N|^2\rangle_\beta,
\end{equation}
where
\begin{equation}
\begin{split}
\label{eq:thermalcontact}
\langle|I_N|^2\rangle_\beta =
Z_N^{-1}\!\!
& \iint\rho_N(\RN,\RN';\beta)\delta(\RNm'-\RNm)\\
&\; \delta(\re_N-\rh_N) \delta(\rep_N-\rhp_N)d\RN\,d\RN'.
\end{split}
\end{equation}
In this equation $Z_N\equiv\mathrm{Tr} \rho_{N}$ is the partition function
for $N$ electron-hole pairs and is needed to normalize $\rho_{N}$ in the 
integral.

The thermal density matrix in Eq.~(\ref{eq:pirate}) may be represented as a
real-space Feynman path integral,\cite{Feynman:1972}
\begin{equation}
\rho(\RN,\RN';\beta) =\int\!\mathcal D \RN(t)\exp\left[-\frac{1}{\hbar}
\int_0^\beta\!\! H dt\right] ,
\end{equation}
where the ends of the paths are $\RN(0)=\RN'$ and $\RN(\beta)=\RN$.
Thus the partition function $Z_N$ and the recombination integral
$\langle |I_N|^2\rangle_\beta$ can be represented by path integrals
that differ only by constraints on the paths,
\begin{align}
\mathrm Z_N&=\int\limits_{\text{diagonal}}
\!\mathcal D \RN(t)\exp\left[-\frac{1}{\hbar}\int_0^\beta\!\! H dt\right]
\\
\mathrm Z_N\langle |I_N|^2\rangle_\beta&=\int\limits_{\text{radiating}}
\!\mathcal D \RN(t)\exp\left[-\frac{1}{\hbar}\int_0^\beta\!\! H dt\right]
\end{align}
The diagonal constraint is the usual trace, $\RN(0)=\RN(\beta)$,
illustrated in Fig.~\ref{fig:path}(a). The radiating constraint is a trace
over the non-radiating pairs, $\RNm(0)=\RNm(\beta)$, and a pairing of
the recombining particles, $\re_N=\rh_N$ and $\rep_N=\rhp_N$,
as illustrated in Fig.~\ref{fig:path}(b).

It is insightful to consider how this path integral formalism for 
the recombination rate, Eq.~(\ref{eq:pirate}), 
relates to the strong and weak confinement 
limits. Consider the $t=0$ slice in imaginary time. In the diagonal 
boundary conditions, the path integral samples the diagonal of the density
matrix in the position basis. For a non-interacting exciton or biexciton,
the electron and hole sample the probability density functions of 
the single particle electron and hole ground states.
In the radiating boundary condition, the electron and hole are forced to 
coincide, but may sample two different points for $t=0_-$ and $t=0_+$. 
The effects approximately cancel out, giving 
$\langle|I_N|^2\rangle_\beta\sim 1$, appropriate for the strong
confinement limit. With the attractive eh-interaction in the weak confinement
limit, the electron and hole pair together in an exciton. In the diagonal 
boundary conditions, the volume sampled by the electron and hole 
is the dot volume $V_{\text{dot}}$ times
the exciton volume $\sim a_X^3$. In the radiating boundary conditions, the 
volume sampled is $V_{\text{dot}}$ for $t=0_-$ times another factor of 
$V_{\text{dot}}$ for $t=0_+$. This gives $\langle|I_N|^2\rangle_\beta\sim 
V_{\text{dot}}/a_X^3$, appropriate for the 
weak confinement limit with dot diameter much less than the wavelength of 
light.

Now consider a bound biexciton. One exciton has radiating boundary conditions 
and the other exciton has diagonal boundary conditions. For the
strong confining case we see a similar cancellation of boundary 
condition effects as for the single exciton.
Since we have contributions from pairing either the spin-up or spin-down 
electrons and holes, we see $\Gamma_{XX}/\Gamma_X\approx 2$. In the weak 
confining limit, the eh-pair in the radiating boundary condition is bound to 
the other eh-pair in the diagonal boundary condition, with a
biexciton radius $a_{XX}$. 
This binding suppresses a factor of $V_{\text{dot}}$
in the biexciton rate, leading to a reduced relative rate, 
$\Gamma_{XX}/\Gamma_X \sim 2 a_{XX}^3/V_{\text{dot}}$. 
While this ratio may drop below one for very large
dots, most self-assembled dots are not much bigger than biexcitons, so we would
not expect to see this limit except in extreme cases.

\section{Computational Methodology}\label{compmethod}

The path integral expression for the recombination rate can be directly 
sampled with Monte Carlo integration, for two- or four-particle interacting
quantum systems.
We have implemented this as a computer simulation that allows for anisotropic
masses and any three-dimensional confining potential we choose.

\subsection{Path integral Monte Carlo}
With the use of Monte Carlo integration,
the path integral approach allows an essentially exact numerical solution 
to many quantum statistical problems.\cite{Ceperley:1995}
Quantum Monte Carlo methods have
been useful for problems related to this one, such as trion binding energies 
in quantum wells,\cite{Bracker:2005,Filinov:2004}
multi-exciton energies in
quantum dots \cite{Shumway:2001a} and positron-electron annihilation 
rates,\cite{Chiesa:2003} as well as bulk phenomena, such as
exciton-exciton scattering \cite{Shumway:2001b,Shumway:2006}
and Bose condensation of excitons.\cite{Shumway:2000b}

To compute $\langle |I_N|^2\rangle_\beta$ we define a
density matrix that contains both radiating \emph{and} diagonal
constraints
\begin{equation}
\tilde{\rho}(\RN,\RN')=\rho_\mathrm{rad}(\RN,\RN')+
\rho_\mathrm{diag}(\RN,\RN')\;,
\end{equation}
where
\begin{equation}
\begin{split}
\rho_\mathrm{rad}(\RN,\RN')=&\rho_N(\RN,\RN')\,
\delta(\RNm-\RNm')\\
&\delta(\re_N-\rh_N)\,
\delta(\rep_N-\rhp_N)
\end{split}
\end{equation}
and
\begin{equation}
\rho_\mathrm{diag}(\bR_N,\bR_N')=\rho_N(\bR_N,\bR_N')\,
\delta(\bR_{N}-\bR_{N}')
\end{equation}
Since the radiating and diagonal constraints form two
disjoint subsets in configuration space we can write the
probability of being in either state as
\begin{equation}\label{eq:pathprob}
\begin{split}
P(\text{radiating/diagonal state})=& \\
\int \tilde{Z}^{-1} \rho_\text{rad/diag}(\bR_N,\bR_N')&\; d\bR_N d\bR_N'
\end{split}
\end{equation}
where $\tilde{Z}=\int \tilde{\rho}(\bR_N,\bR_N')\; d\bR_N d\bR_N'$.
Combining Eqs.~(\ref{eq:thermalcontact}) and (\ref{eq:pathprob}),
we get an expression suitable for evaluation 
within PIMC:
\begin{equation}
\langle |I_N|^2\rangle_\beta=\frac{P(\text{radiating state})}
{P(\text{diagonal state})}\,.
\end{equation}

In our simulations we use a path integral expansion of the density matrix
$\rho_N$ with a finite number of imaginary time slices.
The configuration space of this expansion is $(\bR^{(0)}_N=\bR_N, \bR^{(1)}_N,
\dots, \bR^{(m)}_N=\bR_N')$ where $m$ is the number of time slices. 
We sample the probability distribution $\tilde{Z}^{-1}\tilde{\rho}$ using
the Metropolis algorithm. Since the number of time slices $m$ is of order
$10^4$ in a typical calculation, it is essential to use a
multilevel Metropolis algorithm,\cite{Ceperley:1995} especially when
changing the configuration from radiating to diagonal state and vice versa.
The probability of being in either state
can then be estimated from the relative frequencies $x_\text{rad}$ and
$x_\text{diag}=1-x_\text{rad}$ of radiating
and diagonal path configurations in the Markov chain.

Finally, we arrive at 
$\langle |I_N|^2\rangle_\beta\approx x_{\text{rad}}/x_{\text{diag}}$,
and from Eq.~(\ref{eq:pirate}) we get the radiative recombination rate,
\begin{equation}\label{eq:pimcrate}
\Gamma_N(\beta) = \frac{2nE_{\text{gap}}E_{\text{P}}e^2}{3\hbar^2c^3m}
 \frac{x_{\text{rad}}}{x_{\text{diag}}}.
\end{equation}
When calculating rates, we use the exciton energies from 
the simulation for $E_{\text{gap}}$. 

Since the temperature $k_{\text{B}}T$ in our simulations is small compared to
the single particle level spacing in the dot, we can assume that 
electrons and holes in the biexciton are in a singlet state. Therefore, the
fermion sign problem does not occur in our calculations. (For a review on the
origin of the sign problem, see e.g.~Ref.~\onlinecite{Schmidt:1987}.)

\subsection{Configuration Interaction Calculations}
To demonstrate our method, we have also performed CI calculations on the
same EMA models, Eq.~(\ref{eq:hamil}). The single particle states
are calculated by finite-difference discretization in a cylindrical cell
with 30 nm height and 100 nm diameter, with grid spacing of 0.5 nm
and 0.8 in the vertical and radial directions, respectively; Coulomb
integrals are evaluated by successive over relaxation. This is the same
approach used to calculate multi-exciton states reported in 
Ref.~\onlinecite{Nair:2000}.

For the simulation of excitons and biexcitons in self-assembled QDs, 
our CI expansion uses a 6s5p4d3f2g2h1i basis set including 44 single
particle states. In contrast, the CI expansion by Corni 
{\em et al.}~uses a 4s4p3d basis set (18 single particle states) in
the $xy$-plane and only a single state for the $z$-direction
consisting of Gaussians centered on the dot.
Contrary to the direct expansion of the many-particle wave-function 
in our approach, Corni {\em et al.}~use this basis set first to solve 
the restricted Hartree-Fock
equation to obtain an optimized basis set for the CI expansion. 
Like our approach, the CI calculations by Narvaez {\em et al.}~also use 
single particle states
from a non-interacting Hamiltonian to expand the many-body wave function.
No basis set size is given in Ref.~\onlinecite{Narvaez:2005},
but a previous paper by the same authors using an identical 
method used 6 electron and 10 hole states 
(12 electron and 20 hole states including spin).\cite{Narvaez:2005b}

\section{Tests on parabolic dot}\label{dottest}

To compare our methods, we first consider a model system consisting
of two oppositely charged particles in a harmonic oscillator potential
(``Hooke exciton''). The Hamiltonian then reads
\begin{equation}
H=\sum_{i=1}^{2}\left( \frac{\mathbf{p}_i^2}{2 m_i} +
\frac{m_i \omega}{2} \mathbf{r}_i^2 \right)-
\frac{e^2}{\epsilon \left| \mathbf{r}_1-\mathbf{r}_2 \right|}\;.
\end{equation}
Using center-of-mass and relative coordinates, the problem reduces
to a ordinary differential equation (ODE) that can be integrated numerically
with almost arbitrary exactness.
We apply both the PIMC and CI techniques to this system. Within the
path integral calculations we use a temperature of $\beta=10$ Ha$^*$$^{-1}$,
which is low enough to ensure that only the ground state contributes,
and $m=500$ time slices. The CI calculations use 54 single particle
states to expand the two-particle wave function.

\begin{figure}[tb!]
\psfrag{w [Ha*]}{$\hbar\omega$ (Ha$^*$)}
\psfrag{E [Ha*]}{$E$ (Ha$^*$)}
\psfrag{|I_N|^2}{$\left|I_N\right|^2$}
\centerline{\includegraphics[width=0.68\linewidth]{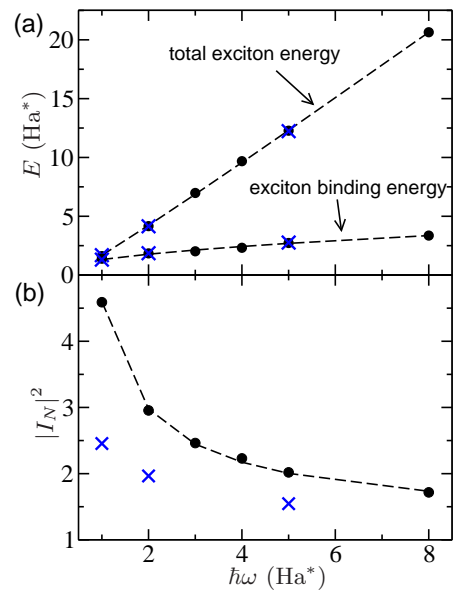}}
\caption{(Color online)
Comparison of results for Hooke exciton from PIMC and CI 
(expansion in 54 basis states): 
(a) total and binding energies and (b) $\left|I_N \right|^2$ 
from integration of ODE, from PIMC and CI (dashed line, circles and 
$\times$'s). Energies are given in units of reduced Hartrees
Ha$^*= \mu e^4/\hbar^2 \epsilon^2$, where
$\mu$ is the reduced mass of the exciton. Error bars 
are of the order of symbol size.} 
\label{pimcvsci}
\end{figure}

Figure \ref{pimcvsci}(a) shows the total and the binding energy of the two
particles. Both CI and PIMC show very good agreement with the results from 
numerical integration of the ODE. 
However, for $\left| I_N\right|^2$ only PIMC shows good convergence
whereas the CI results are in general too low, up to a factor of 2
in the weak confinement case. But even for strong confinement there is a 
considerable discrepancy, although the confinement energy significantly 
exceeds the exciton binding energy (see Fig.~\ref{pimcvsci}(b)). 
In our calculations we have also found that the CI result for 
$\left| I_N\right|^2$ approaches the correct value rather slowly 
with increasing basis set size, thus leading to a false impression
of convergence. If only the dependence of the result on the basis set is 
used as a measure of convergence, it is hard to decide whether 
a calculation has converged or not.

The true many particle wave function for Coulombic interactions must have
a coalescence cusp for $\mathbf{r}_1=\mathbf{r}_2$,\cite{Kato:1957}
but a CI expansion of the wave function in products of smooth single particle
basis functions cannot have a non-analytic behaviour. 
Convergence problems of CI associated with the failure to
reproduce this cusp are for example solved
by using correlated basis functions 
(e.g.~Ref.~\onlinecite{kutzelnigg:1991,%
termath:1991,klopper:1991}). 
It is therefore not surprising that CI calculations 
give better results for energies than
for the overlap matrix element: The energy is calculated using
the wave function at every grid point, whereas for the decay rate
mainly the cusp at $\mathbf{r}_1=\mathbf{r}_2$ enters. 
That the overlap matrix element is more sensitive to errors in the 
wave function
than the energy, also shows in the fact that even a Hartree-Fock calculation
gets up to 95\% of the exciton binding 
energy\cite{Shumway:2001a} but completely lacks the correlation cusp, 
leading to a decay rate that does not depend on the dot size.\cite{Corni:2003}
Still, it is striking that the CI results are able to reproduce
energies very accurately
and yet completely fail to obtain the correct overlap matrix element.
PIMC does not suffer from a finite basis set and can thus reproduce both
energies and the overlap matrix element very accurately.

\section{Results for self-assembled dots}\label{results}

\begin{table*}
\caption{\label{table:parameters}
Parameters used in the calculations.}
\begin{ruledtabular}
\begin{tabular}{cccccccccc}
dot/barrier& $E_{\text{gap}}^{\text{barrier}} (eV)$ & $\epsilon$
& $m_e$ & $m_h^\parallel$ & $m_h^\perp$
& $\Delta V_e$ (eV) & $\Delta V_h$ (eV) & $t_{\text{WL}}$ (\AA) & 
$E_{\text{P}}$ (eV)\\
\hline
In$_{0.5}$Ga$_{0.5}$As/GaAs &
1.519\footnotemark[1] \footnotetext[1]{Ref.~\onlinecite{Landolt:1982iiia}.}&
12.5\footnotemark[2] \footnotetext[2]{We approximate the strained InGaAs 
material in the dot by just taking the bulk GaAs 
value.\cite{Landolt:1982iiia}}&
0.067\footnotemark[2]&
0.11\footnotemark[2]&
0.38\footnotemark[2]&
0.250\footnotemark[3]\footnotetext[3]{Estimated from strain-modified band 
offsets plotted in Ref.~\onlinecite{Shumway:2001c}.}&
0.200\footnotemark[3]&
16, 0&
25.7\footnotemark[2]\\
InP/In$_{0.5}$Ga$_{0.5}$P &
 1.920\footnotemark[4]\footnotetext[4]{Estimated from EPM/VFF calculations 
(Nair, unpublished).}&
 12.6\footnotemark[5]\footnotetext[5]{Bulk InP value.\cite{Landolt:1982iiia}}&
 0.079\footnotemark[5]&
 0.150\footnotemark[5]&
 0.600\footnotemark[5]&
 0.420\footnotemark[4]&
 0.070\footnotemark[4]&
 5&
 20.4\footnotemark[5]\\
CdSe/ZnSe &
 2.820\footnotemark[1]&
\phantom{1}9.3\footnotemark[6]\footnotetext[6]{Bulk CdSe 
values.\cite{Landolt:1982iiia}}&
 0.130\footnotemark[6]&
 0.380\footnotemark[7]\footnotetext[7]{Ref.~\onlinecite{Ekimov:1992}.}&
 1.000\footnotemark[7]&
 0.735\footnotemark[8]\footnotetext[8]{CdSe/ZnSe band offsets chosen 
to match simulations in Ref.~\onlinecite{Puls:1999}.}&
 0.135\footnotemark[8]&
 5&
 17.5\footnotemark[7]\\
\end{tabular}
\end{ruledtabular}
\end{table*}

We have applied these techniques to common single-band effective
mass models of quantum dots, summarized in Table \ref{table:parameters}.
We chose these materials and sizes because of availability of published
experimental values. The dot geometry is a lens shape, with a height to
diameter ratio of 1:10. The calculations include
a wetting layer, modeled as a quantum well with thickness $t_{\text{WL}}$ 
extending from the base of the dot. The dot potential consists of potential 
steps of finite height $V_{e}$ ($V_{h}$) 
for electrons (holes) at the boundaries of the lens and the wetting layer. 
The three systems we have studied are:
\begin{enumerate}
\item {\em InGaAs/GaAs:}
This is the most studied material for optical properties of self-assembled
dots, and we are comparing our results with four separate PL rate
experiments. Some of these dots
are grown as alloyed InGaAs material, while others are nominally pure InAs.
Even for nominally pure dots, intermixing and annealing at high temperatures 
often leads to dots with significant Ga content. Based on reported
growth conditions and 
PL energies, we have chosen to simulate dots composed of In$_{0.5}Ga_{0.5}As$.
The dot diameters, from 10 nm to 60 nm, cover the size range for nearly all 
dots of this material reported in PL studies.
We have included a 6 monolayer (ML), or 16 \AA, In$_{0.5}$Ga$_{0.5}$As wetting
layer under the dot.\cite{Wojs:1996}
To show the influence of the wetting layer we also give results for 
$t_{\text{WL}}=0$.
\item {\em InP/InGaP:}
We have included a 2 ML, or 5 \AA, InP wetting layer under the dot.
\item {\em CdSe/ZnSe:}
We have included a 2 ML, or 5 \AA, CdSe wetting layer under the 
dot.\cite{Lee:1998c}
\end{enumerate}

While our path integral formalism allows for a thermal distribution of
initial states, we have chosen a low temperature ($T \approx 8K$) so
that we consider only emission from the ground state. We have
discretized imaginary time in the path integral in steps of 
$\tau= 1.3\times 10^{-5} K^{-1}$.
The simulation time for one dot diameter was approx. 200 min for the exciton 
and 350 min for the biexciton on 10 Athlon MP 1600+ processors. 

\begin{figure*}
\centerline{\includegraphics[width=0.96\linewidth]{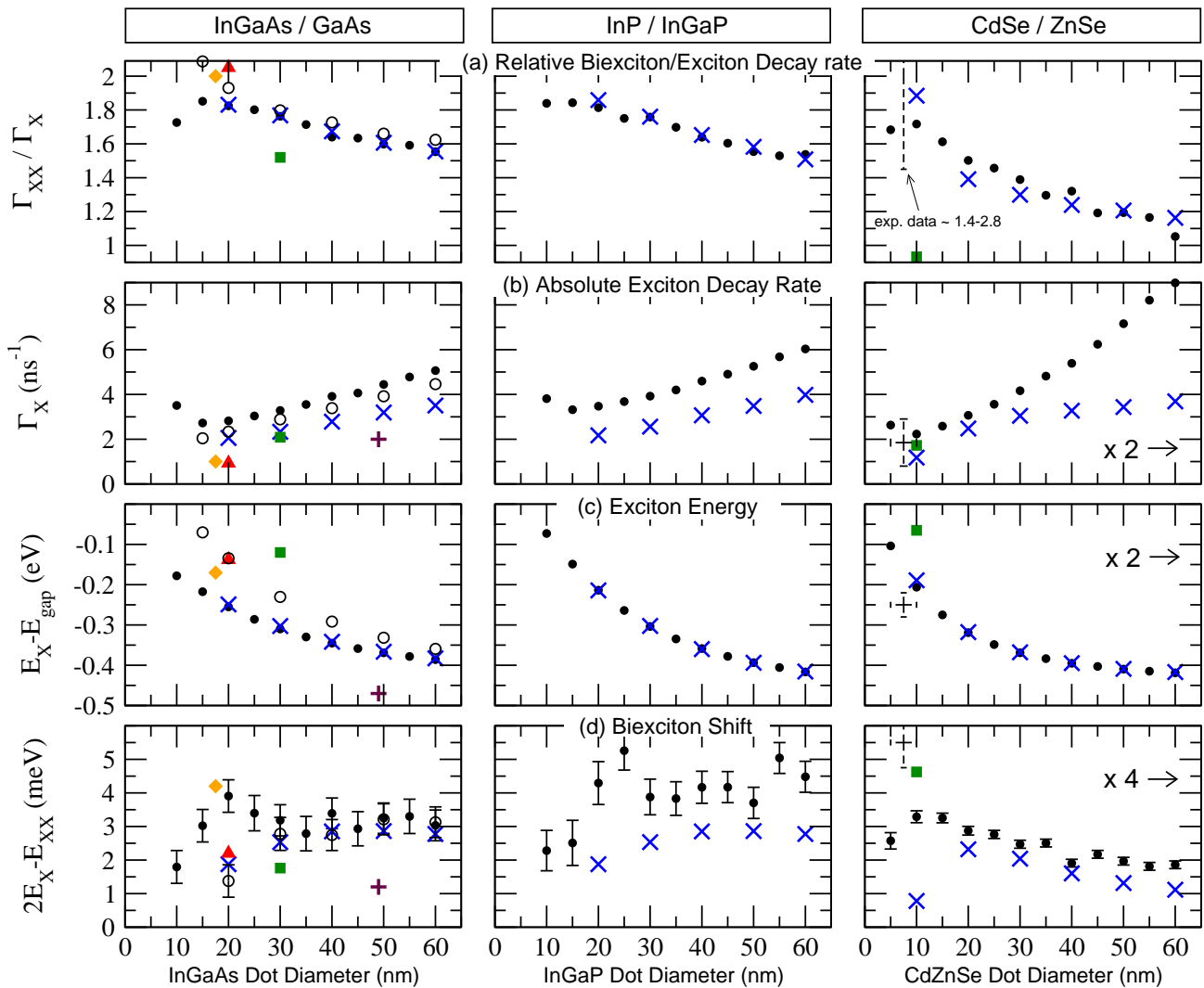}}
\caption{(Color online)
Our results of path-integral calculations for for
InGaAs/GaAs, InP/InGaP, and CdSe/ZnSe lens-shaped dots
of different diameters, with height/diameter = 0.1.
Rows of panels present (a) the relative decay rate of biexciton to exciton,
(b) the absolute decay rate of exciton, (c) the exciton energy, at which
the exciton luminescence peak would be observed,
and (d) the energy shift of the biexciton luminescence peak.
Solid circles are path integral results for $t_{\text{WL}}>0$, open circles
for $t_{\text{WL}}=0$. Error bars are only given if the error exceeds the 
symbol size. 
We compare to our configuration interaction calculations
for $t_{\text{WL}}>0$ ($\times$'s),
which miss some of the correlation. We also show experimental data
for InGaAs/GaAs ($\triangle$'s are data from Ref.~\onlinecite{Thompson:2001}, 
where we have assumed $d=20$ and used their measured $X^*$ and $XX^*$
data points;
$\Box$'s are data from Ref.~\onlinecite{Santori:2002}, $\Diamond$'s are from
Ref.~\onlinecite{Ulrich:2005}, and $+$'s are from Ref.~\onlinecite{Kono:2005})
and for CdSe ($\Box$'s are data from Ref.~\onlinecite{Bacher:1999},
the range of data from 
Ref.~\onlinecite{Patton:2003} is indicated by dashed crosses).
\label{fig:results}}
\end{figure*}

In Fig.~\ref{fig:results} we present results of our path integral calculations,
along with our CI results and published experimental data points.
In Fig.~\ref{fig:results}(b) we see that the absolute exciton decay rate
$\Gamma_X$ increases for large dots with increasing dot diameter due to 
the larger exciton coherence volume. As already expected from our model 
calculations, the CI results for the decay rate suffer from underconvergence, 
although the exciton energies from PIMC and CI agree very well. This is 
particularly evident in the decay rate for CdSe/ZnSe where the CI result begins
to saturate for large dot sizes due to missing correlation, 
whereas the Monte Carlo result still increases uniformly with
increasing dot diameter. A similar flattening of the decay ratio
with increasing dot diameter can also be observed in the CI results
of Corni {\em et al.}, possibly indicating missing correlation at larger dot 
sizes.
For small dots, we observe a minimum of the decay rate 
when the dot height $h$ becomes comparable to the wetting layer 
thickness $t_{\text{WL}}$. 
The InGaAs/GaAs dots without wetting layer do not show this 
behaviour. Note that in this case we only give results 
down to dot diameter $d=15$ nm, because the exciton becomes unbound for smaller
dot sizes. The exciton decay rate in the InGaAs/GaAs material system is
larger for $t_{\text{WL}}=16$ \AA{} than for $t_{\text{WL}}=0$ \AA{} since 
the effective dot size and thus the exciton coherence volume is larger for the
dots including a wetting layer. 

The relative decay rate $\Gamma_{XX}/ \Gamma_{X}$ of the biexciton,
Fig.~\ref{fig:results}(a), varies from
approx.~2 down to 1.5 for InGaAs/GaAs and InP/InGaP and even down to 1 for the
CdSe/ZnSe material system. For large dots we observe a decrease of
$\Gamma_{XX}/ \Gamma_{X}$ for increasing dot size, at small dot sizes there is
a maximum of the 
relative biexciton decay rate corresponding to the minimum in the exciton
decay rate. Again, the data for InGaAs/GaAs without wetting layer does not
show an extremum for small dots but reaches
$\Gamma_{XX}/ \Gamma_{X} \approx 2$, corresponding
to the strong confinement or non-interacting limit. As we explained above, CI
underestimates decay rates. However, since it underestimates both exciton and
biexciton decay rates, the relative ratio from CI is actually rather similar to
the Monte Carlo result.

\begin{figure}
\includegraphics[width=0.85\linewidth]{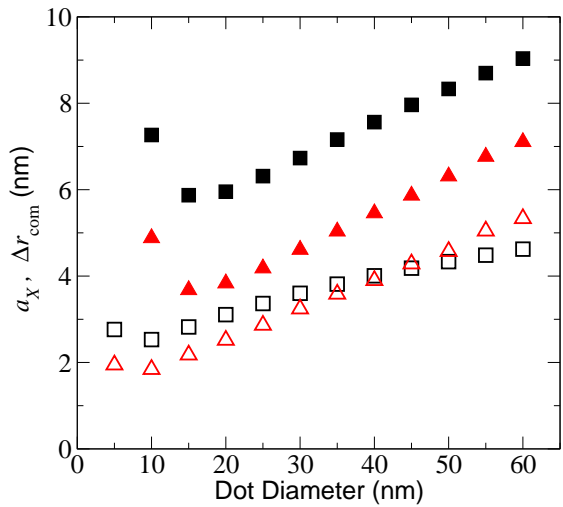}
\caption{(Color online)
Exciton radius $a_X$ ($\Box$'s) and 
exciton center of mass fluctuations $\Delta r_{\text{com}}$ 
($\triangle$'s) for InGaAs/GaAs dots with a 6 ML 
wetting layer (full symbols) and for CdSe/ZnSe dots (open symbols).}
\label{fig:spatial}
\end{figure}

To gain more insight into the size-dependence of the decay rates, it is useful 
to study the spatial extent of the exciton wave function in the dot. 
The decay rate is closely linked to the coherence volume and thus to the 
volume that is filled by the exciton wave function. In 
Fig.~\ref{fig:spatial} we present the size-dependence of the exciton radius
$a_{X}$ and the standard deviation of the exciton center of mass (com) 
coordinate $\Delta r_{\mathrm{com}}=\sqrt{\langle \mathbf{r}_{\text{com}}^2 \rangle-
\langle \mathbf{r}_{\text{com}} \rangle^2}$ for the InGaAs/GaAs 
and CdSe/ZnSe material systems. The results for InP/InGaP quantum dots
are similar to those for InGaAs/GaAs, just as in Fig.~\ref{fig:results}. 

For small dot sizes we find a minimum of both $a_{X}$ and 
$\Delta r_{\mathrm{com}}$
corresponding to the minimum in the decay rate. As the dot height becomes 
comparable to the wetting layer thickness, the exciton center of mass moves 
into the wetting layer and the wave function extends further
into the quantum well underneath the dot. A dot height less than 
$t_{\text{WL}}$ thus corresponds to an effectively larger dot size. The 
increased coherence volume leads to an increase in the decay ratio 
$\Gamma_{X}$ and a decrease of the relative biexciton ratio 
$\Gamma_{XX}/ \Gamma_{X}$. In the limit of zero dot size, 
we would be in the quantum well situation, however, the dipole approximation
leading to Eq.~(\ref{eq:envelopedipole}) does not hold for an extended quantum 
well state.\cite{Sugawara:1995} In the case 
of the QD without wetting layer the coherence volume decreases 
monotonically until the exciton becomes unbound, thus we do not observe 
an extremum in the decay rates. 

In the case of larger dot sizes we observe different behaviour for InGaAs/GaAs
and CdSe/ZnSe: In the case of InGaAs/GaAs, $\Delta r_{\mathrm{com}}<a_{X}$ 
for all studied diameters and the exciton radius $a_{X}$ does not
show saturation towards the bulk value. With respect to these properties of the
wave function, the dots are in the strong confinement limit, whereas
the decay rate shows signatures of strong to intermediate
confinement: The relative biexciton ratio can be tuned over a 
relatively large range---from 2 to 1.5---by changing the dot geometry, 
an effect entirely due to electronic correlation.
  
For CdSe/ZnSe we find a crossover to 
$\Delta r_{\mathrm{com}}>a_{X}$ with increasing dot diameter. This corresponds 
to the weak confinement limit---an exciton ``bouncing'' around in the 
dot---and thus we observe relative biexciton ratios down to 1.
However, reported photoluminescence measurements on this type 
of QDs are usually carried out on dots with a diameter $d$ around 10 nm,
so that the 
relevant experimental data for CdSe/ZnSe also lies in the strong to 
intermediate confinement regime.
  
When comparing to experiment, we notice that for some of the reported
data our calculated exciton energies are much smaller than the experimental
values. In these cases (Refs.~\onlinecite{Santori:2002},%
\onlinecite{Bacher:1999}) the growth conditions enhanced alloying. Our
parameters for the band offsets do not seem to describe these shallow dots very
well. However, since we use an abrupt potential step for the QD boundaries, 
the step height should not influence our calculated decay 
rates significantly, as long 
as the exciton is still bound. The step height just determines the 
exponential decay length of the wave function into the barrier, so its 
influence on the wave function inside the dot is only indirect. 
Comparison of our calculated results with these experiments is thus 
still valid.

In comparing to the work of Thompson {\em et al.},\cite{Thompson:2001}
we found it was necessary to re-identify their reported exciton spectra line
as a charged exciton. Our concerns were their reported negative (blue-shifted)
biexciton binding energy and their ratio $\Gamma_{XX}/\Gamma_X\approx2.3$.
Their spectra are very similar to spectra reported by Lomascolo {\em et al.},
which are dominated by charged exciton,\cite{Lomascolo:2002}
only with the exciton/charged exciton labels switched.
Since Thompson {\em et al.}~called their identification of the charged and
neutral exciton tentative, and did not offer any alternative explanation 
for the unusual energy shift and relative decay rate of their supposed neutral
exciton/biexciton pair, we chose to compare to the data they had attributed to
the charged exciton and charged biexciton. These states have
$\Gamma_{XX}/\Gamma_X\approx 2.06$ and a biexciton binding energy
of $+2$ meV.
This identification makes little
difference in our comparison of decay rates, but does give us much
better agreement for the positive biexciton binding energy and is
consistent with the relative biexciton decay rate in the strong-confinement 
limit. Thompson {\em et al.}~do not give any value for the dot size in
their experiment, but claim that their dots are smaller than those
of other photoluminescence experiments. 
We chose to attribute their data a dot diameter of 20 nm,
based on our energy calculations for the InGaAs/GaAs dot without wetting layer.
In the absence of further information about the wetting layer
thickness in the experiment, this seems 
reasonable since the nominal InAs coverage in this experiment is only 1.7ML. 
Accordingly, we will always compare the experimental data from 
Thompson {\em et al.}~as well as the data from 
Ulrich {\em et al.}\cite{Ulrich:2005}~(experimental 
wetting layer thickness 1 ML) with the results of our 
calculation with $t_{\text{WL}}=0$ ML.

Our PIMC results for the exciton decay rate in InGaAs/GaAs 
agree with experiment within about a factor of two, but seem to systematically
overestimate the decay rate. This could be due to our 
simplified model of an ideal dot. The agreement with experiment could,
for example, be 
improved by including the effects of alloying in the dot potential.
Disorder introduced by alloying leads to stronger localization of the particles
in the dot,\cite{Klimeck:2002} thus reducing the coherence volume and 
the electron-hole overlap. Another possiblity for improving the path integral
results would be to use a dot potential from strain 
calculations,\cite{Shumway:2001c}
since strain might also lead to an increased 
electron-hole separation.\cite{Lee:2000}
The inclusion of such single-particle potentials in
PIMC is perfectly feasible and does not introduce any additional 
computational cost. Even in the strong confinement or non-interacting
limit $\left|I_{N}\right|^2 \approx 1$, so 
$\Gamma_{X}\gtrapprox 2\;\mathrm{ns}^{-1}$ using the parameters from Table 
\ref{table:parameters}. Thus the low decay rate from 
Refs.~\onlinecite{Thompson:2001} and \onlinecite{Ulrich:2005}
cannot be explained in our model, hinting at the need of a more detailed
dot potential. However, for the study of the size dependence 
of the decay rates a model potential is perfectly valid and 
yields results that are easier to interpret. 

The Monte Carlo calculations can reproduce the range of observed relative
biexciton ratios from 2 to 1.5. The data from Refs.~\onlinecite{Thompson:2001}
and \onlinecite{Ulrich:2005} is described very well by the QD without 
wetting layer, whereas the data from Ref.~\onlinecite{Santori:2002} 
seems to be best reproduced by a QD
with wetting layer, although we have to assume a somewhat larger 
effective dot size. The calculated biexciton binding energies are also
close to the experimental values. The extremely low relative biexciton
decay rate from Ref.~\onlinecite{Kono:2005}, $\Gamma_{XX}/ 
\Gamma_{X}=0.33$ however cannot be explained at all in our model. In the 
original paper, the low biexciton decay rate was attributed to weak 
confinement effects, but from our calculations
we can conclude that InGaAs/GaAs QDs with diameters around 50 nm are still far 
from the weak confinement regime. 

We are not aware of any studies on the exciton and biexciton dynamics
in single InP/InGaP QDs, but our calculations are within
the reported exciton life time range 
of 100--500 ps for QD ensembles with dot
diameters between 20 nm and 40 nm.\cite{Hatami:2003,Hatami:2000}

When comparing to the experimental data on single CdSe/ZnSe dots by
Patton {\em et al.},\cite{Patton:2003} we chose to only give the range
of the reported data. The variation of exciton energies in
Patton {\em et al.}~is attributed to different localization potentials, 
i.e.~different Cd concentration, 
and not to different dot sizes. The dot diameter 
for their samples is reported to be between 5 nm and 10 nm.\cite{Gindele:1999}
The CdSe/ZnSe exciton decay rates calculated by PIMC agree very well with the 
reported experimental data. However, we completely fail to reproduce the 
relative biexciton ratio $\Gamma_{XX}/ \Gamma_{X}\approx 1$ from 
Bacher {\em et al.}\cite{Bacher:1999} Such a low
biexciton decay rate is only to be expected for very large dots 
in our simulation. The experiments by Patton {\em et al.}~on 
very similar sized QDs in contrast yielded a relative biexciton ratio 
$\Gamma_{XX}/ \Gamma_{X}\approx 2$ with a rather large experimental
spread ($\Gamma_{XX}/ \Gamma_{X}\approx 1.4-2.8$). We expect CdSe/ZnSe QDs of
about 10 nm to be towards the strong confinement limit, consistent 
with the experiment by Patton {\em et al.} Therefore we presume that more
knowledge about the dot potential would be needed to explain
the results of Bacher {\em et al.}, a simple box model, as suggested in 
Ref.~\onlinecite{Bacher:1999}, is certainly not enough. The fact that 
the exciton energies from Ref.~\onlinecite{Patton:2003} are well explained 
by our model whereas the rather shallow dots from Ref.~\onlinecite{Bacher:1999}
are not, supports this presumption. 
It should also be noted that the QDs of Ref.~\onlinecite{Ulrich:2003}
were grown under conditions similar to those of Bacher 
{\em et al.} They show recombination
energies and exciton lifetimes comparable to the results of Bacher 
{\em et al.}, but much shorter biexciton lifetimes. 
Since no estimate of the dot 
size was given, we cannot directly
compare to our calculations, but the reported ratio of 
$\Gamma_{XX}/ \Gamma_{X}\approx 1.4$ agrees
well with the ratios expected from our calculation.

We cannot compare our hitherto obtained rates with the calculations of Corni
{\em et al.}~on the size-dependence of the exciton and biexciton decay rates
because of the different dot potentials. However, if we apply our technique to
the truncated parabola potentials used in their study, we obtain exciton 
decay rates that are for small dots around 50\%, and for large dots 
even up to two times larger than the results of Ref.~\onlinecite{Corni:2003}. 
Given that even our CI
expansion, using a large basis set of 44 single particle states, 
yields absolute rates that are 
too low, it is not surprising that the much smaller basis set of 
Corni {\em et al.}~also fails to calculate absolute rates. 
Yet, the CI calculations show the right trends---increasing 
decay rate and decreasing $\Gamma_{XX}/ \Gamma_{X}$ 
with increasing dot size---compatible 
with our results. Also, the relative biexciton decay rate from 
PIMC is very similar to the one obtained by Corni {\em et al.}

Narvaez {\em et al.}~have performed CI calculations on the height dependence of
recombination rates in lens-shaped In$_{0.6}$Ga$_{0.4}$As/GaAs quantum 
dots with a fixed diameter (25.2 nm), using atomic 
pseudopotentials and a realistic model for alloying. Their calculates decay 
rates lie in the range of 0.4--0.5 ns$^{-1}$, a factor of four lower than 
our results, but also a factor of two lower than experimental 
values.\cite{Thompson:2001,Ulrich:2005} Their 
reported relative biexciton decay ratio $\Gamma_{XX}/ \Gamma_{X}=4$ is a 
factor of 2 larger than what is expected for strong confinement and is to our 
knowledge not observed in experiment. 
Path integral techniques cannot be adapted to using pseudopotentials 
easily, and thus we cannot directly 
compare results. Still, from our experience with the calculation of rates 
from CI, we are somewhat concerned with the absolute value of the rates. 
The reported basis set
size in Ref.~\onlinecite{Narvaez:2005b} is much smaller than the one used in 
this study. For example, their minimum of the exciton lifetime 
(corresponding to a maximum in the decay rate) at a dot height
of 65\AA{} could possibly be a sign of missed correlation at larger dot sizes: 
From our calculations we would expect the decay rate to grow monotonically 
with dot volume. The decrease of the exciton lifetime found at 
smaller dot heights however is 
compatible with our findings.  

\section{Conclusion}\label{conclusions}

We have developed a path integral Monte Carlo approach for studying 
exciton and biexciton recombination rates in self-assembled quantum
dots. This technique allows us to study general 3-d potentials for
a wide variety of single-band EMA models. Our calculations indicate
that self-assembled dots are in the strong to intermediate confinement
regime, where Coulomb correlation effects are becoming important. 
In particular, for large dots we see a clear monotonic rise 
in recombination rate versus diameter, and a decrease in the 
relative biexciton decay rate, $\Gamma_{XX}/\Gamma_X$. 
From our calculations we can state that relative
decay rates $\Gamma_{XX}/\Gamma_{X}\approx 1.5-2$ are expected
for typical photoluminescence experiments, an effect due entirely 
to correlation.
We have seen that quantum dots of the size used in PL experiments tend towards
the strong confinement regime. Thus, the low relative biexciton decay rates
$\Gamma_{XX}/\Gamma_{X}\leq1$ from some 
experiments,\cite{Kono:2005, Bacher:1999}
that were attributed to weak confinement, cannot be explained 
by weak confinement effects.
It should be noted that in single dot experiments rather large
dot-to-dot fluctuations have been reported.\cite{Patton:2003}
Given the spread of experimental data, our calculations compare rather 
favorably against experiment. 

We have further shown that CI expansions using uncorrelated 
single particle basis sets have severe shortcomings in calculating
decay rates. Rather surprisingly, we have found that CI expansion 
in a large basis set of 44 states underestimate decay rates by far,
even for dot sizes comparable to the exciton Bohr radius
and although the calculated energies were well-converged. 
Yet, due to a cancellation of errors, the relative biexciton 
decay rate calculated by CI was found to be similar to the path integral 
result. Also, trends were in general reproduced correctly by CI. 
CI has some advantages over 
PIMC, such as being able to use an atomic description of the quantum dot. 
However, absolute decay rates from CI must be regarded with caution.

In conclusion, we have developed a microscopic path-integral technique for 
calculating exciton and biexciton decay rates, that fully 
treats quantum correlation in realistic models.
We can apply this to arbitrary geometries within single-band EMA models.
Our calculations on lens-shaped self-assembled dots indicate that
these commonly studied structures are in the regime between
strong and intermediate confinement.
The formalism has a built-in thermal distribution of carriers that we have
not yet exploited. Another area for future research is an extension of 
this technique to semiconductors with indirect band gaps, such as Si/Ge.

\begin{acknowledgments}
MW would like to thank J. Kainz and M. Sch\"utz for useful discussions and
acknowledges financial support from DFG within GRK 638.
Work by JS supported by NSF DMR-0239819.
\end{acknowledgments}

\bibliographystyle{apsrev}
\bibliography{dots}

\end{document}